\documentclass[12pt]{article} 
\usepackage{a4wide,amssymb,cite,axodraw,float,cite} 
\usepackage{latexsym,epsfig,graphicx}

\def\de{\partial} 
\def\a{\alpha} 
\def\b{\beta}

\def\d{\delta}

\def\la{\lambda} 
\def\La{\Lambda} 
\def\k{\kappa} 
\def\m{\mu} 
\def\n{\nu} 
\def\r{\rho} 
 
\def\s{\sigma}

\def\th{\theta} 
 
\newcommand{\be}{\begin{equation}} 
\newcommand{\ee}{\end{equation}} 
\newcommand{\bea}{\begin{eqnarray}} 
\newcommand{\eea}{\end{eqnarray}} 
\newcommand{\beqar}{\begin{eqnarray*}} 
\newcommand{\eeqar}{\end{eqnarray*}} 
\newcommand{\eg}{{\it e.g.,}\ } 
\newcommand{\ie}{{\it i.e.,}\ } 
\newcommand{\reef}[1]{(\ref{#1})} 
\newcommand{\nn}{\nonumber} 
 
\begin{document}

\begin{titlepage} 

\rightline{hep-th/0507278} 
\rightline{July 2005}

\begin{centering} 
\vspace{1cm} 
{\large {\bf Cosmological Evolution of a Purely Conical \\ Codimension-2 Brane World }}\\ 

\vspace{1.5cm}

 {\bf Eleftherios~Papantonopoulos}$^{a,*}$ and {\bf Antonios~Papazoglou} $^{b,**}$ \\ 
\vspace{.2in} 
 
$^{a}$ National Technical University of Athens,\\ Physics 
Department,\\ Zografou Campus, GR 157 80, Athens, Greece. \\ 
\vspace{3mm} 
$^{b}$ \'Ecole Polytechnique F\'ed\'erale de Lausanne,\\ Institute of Theoretical Physics,\\ 
SB ITP LPPC BSP 720, CH 1015, Lausanne, Switzerland. 
 
\end{centering} 
\vspace{2cm} 
 
\begin{abstract}

We study the cosmological evolution of isotropic matter on an 
infinitely thin conical codimension-two brane-world. Our analysis 
is based on the boundary dynamics of a six-dimensional model  in 
the presence of an induced gravity term on the brane and a 
Gauss-Bonnet term in the bulk. With the assumption that the bulk 
contains only a cosmological constant $\La_B$, we find that the 
isotropic evolution of the brane-universe imposes a tuned relation 
between the energy density and the brane equation of state. 
The evolution of the system has fixed points (attractors), which 
correspond to a final state of radiation for $\La_B=0$ and to de 
Sitter state for $\La_B>0$. Furthermore, considering anisotropic 
matter on the brane, the tuning of the parameters is lifted, and 
new regions of the parametric space are available for the 
cosmological evolution of the brane-universe. The analysis of the 
dynamics of the system shows that, the isotropic fixed points 
remain attractors of the system, and for values of $\La_B$ which 
give acceptable cosmological evolution of the equation of state, 
the line of isotropic tuning is a very weak attractor. The initial 
conditions, in  this case, need to be fine tuned to have an 
evolution with acceptably small anisotropy.

\end{abstract}

\vspace{1cm} 
\begin{flushleft} 
 
$^{*}~$ e-mail address: lpapa@central.ntua.gr \\ 
$ ^{**}$ e-mail address: antonios.papazoglou@epfl.ch

\end{flushleft} 
\end{titlepage}

\section{Introduction}

Cosmology in theories with branes 
 embedded in  extra dimensions has been the subject of intense investigation during the last years. 
  The most detailed analysis has been done for brane world models in five-dimensional space \cite{5d}. 
   The effect of the extra dimension can modify the cosmological evolution, depending on the model, 
    both at early  and late times. Both modifications can be interesting phenomenologically. 
     For example, the early time modification may give less fine tuned inflationary parameters 
     \cite{Maartens:1999hf} and the  late  time modifications may shed light to the recent 
      acceleration of the universe \cite{Deffayet}.

Much less has been done, however, for the cosmology of theories in 
six or higher dimensions with branes of codimension greater than 
one. This is because, unlike the codimension one case, these 
branes exhibit bulk curvature singularities which are worse than 
$\d$-function singularities. They then need some regularization 
(introduction of brane thickness) which makes the study of 
cosmology on them rather complicated \cite{thickcosmo}. An 
alternative way to study cosmologies of branes of higher 
codimension would be  to consider corrections to the gravitational 
action, such as an induced curvature term on the brane 
\cite{induced} and a Gauss-Bonnet term in the bulk \cite{GB}, 
which allow the brane to have a mild singularity structure (see 
also \cite{intersecting}). These thin brane cosmologies would have 
the additional advantage that the internal structure of the brane 
does not influence the macroscopic cosmological evolution.

The case of codimension two is particularly interesting because 
these branes have a special property. Their vacuum energy 
(tension) does not curve their world-volume  but only induces a 
deficit angle on the branes \cite{Chen:2000at}. There have been 
several attempts to utilize this property in order to self tune 
the effective cosmological constant  to zero and provide a 
solution to the cosmological constant problem \cite{6d}. The 
requirement of simple singularity structure in this case means 
that the branes are purely conical. If higher dimensional gravity 
is conventional (\ie if there is only a higher dimensional 
Einstein term in the action), it has been found that cosmological 
evolution on the brane is not possible, because only pure tension 
is allowed on the brane \cite{Cline:2003ak}. The way out of this 
problem, keeping the structure of the singularity simple, is, as 
mentioned before,  to assume more complicated higher dimensional 
gravity dynamics, as \eg the introduction of a Gauss-Bonnet term 
in the bulk \cite{Bostock:2003cv} and an induced gravity term on 
the brane \cite{Janpaper}. 
 
The latter modifications, have as a consequence the existence of a 
brane Einstein equation which is purely four dimensional 
\cite{Bostock:2003cv,Janpaper} with the mere addition of a deficit 
angle dependent cosmological term. In our previous work 
\cite{Janpaper}, we have shown that apart from this Einstein 
equation, the bulk Einstein equations evaluated on the brane 
provide a constraint on the matter evolution on the brane. If only 
an induced gravity term is added, this constraint corresponds to a 
tuning between brane and bulk matter. If in addition a 
Gauss-Bonnet term is included in the bulk, this constraint can 
either be  a matter tuning or a dynamical equation depending on 
the symmetries of the spacetime metric. This happens because the 
constraint involves the Riemann tensor of the induced four 
dimensional metric. If the induced metric is isotropic, the 
Riemann tensor can be expressed in terms of the Ricci tensor and 
the scalar curvature and, through the brane Einstein equation, 
will give a tuning between the matter of the bulk and the brane. 
If the four-dimensional metric is not isotropic, the Riemann 
tensor is independent of the other two known curvature quantities 
and the constraint gives rise to a dynamical equation for the 
anisotropy. 
 
In the present paper, we will study in details the cosmological 
evolution of a conical brane  with both  an induced gravity and a 
Gauss-Bonnet term added in the higher dimensional gravity action. 
For simplicity we will assume that the only matter in the bulk is 
cosmological constant $\La_B$. We will solve the equations of 
motion evaluated on the brane and assume that the integration 
of them in the bulk does not give rise to pathologies (\eg 
singularities).

Firstly, we will study the isotropic cosmology, in which the brane 
matter has to obey a tuning relation. We will see that the 
evolution of the system for  $\La_B=0$ tends to a fixed point with 
$w=1/3$ and  for $\La_B>0$ to a fixed point with $w=-1$. For 
$\La_B<0$ the system has a runaway behaviour to $w \to \infty$. 
 
We will then relax the isotropy requirement for the metric 
(keeping, however, the matter distribution isotropic) in order to 
find whether the above matter tuning is an attractor or not. The 
matter on the brane need not now satisfy the previous tuning 
relation and the allowed regions of  initial values of the energy 
density and pressure are significantly larger.  The analysis of 
the dynamics of the system shows that line of isotropic tuning is 
an attractor for $\La_B \geq 0$ and thus the system isotropises 
towards it. However, for values of $\La_B$ which give a realistic 
cosmological evolution of the equation of state, the attractor 
property of the previous line is very weak and fine tuning of the 
initial conditions is necessary in order to have an evolution with 
acceptably small anisotropy. For $\La_B < 0$ the system shows, as 
in the isotropic case, a runaway behaviour.  We will discuss in 
detail all the above cases and finally present our conclusions.

\section{Boundary Einstein Equations for a Conical Codi-mension-2 Brane-world Model} 
 
We consider a six-dimensional theory with general bulk dynamics 
encoded in a Lagrangian ${\cal L}_{Bulk}$ and a 3-brane at some 
point $r=0$ of the two-dimensional internal space with general 
dynamics ${\cal L}_{brane}$ 
 in its world-volume. The gravitational dynamics is described by a Gauss-Bonnet term 
 in the bulk and an induced four-dimensional curvature term localized at the position of the 
 brane. Then the total action is written as 
\bea 
 {\cal S}&=&{M^4_6 \over 2}\left\{ \int d^6x \sqrt{-g^{(6)}}[R^{(6)}+\a(R^{(6)~2}-4R^{(6)~2}_{MN}+R^{(6)~2}_{MNK\Lambda})] 
 +r_c^2\int d^6x 
\sqrt{-g}R^{(4)} ~{\d(r) \over 2 \pi L} \right\} \nn \\  &+& \int 
d^6x {\cal L}_{Bulk} + \int d^6x {\cal L}_{brane}~{\d(r) \over 2 
\pi L}~,\label{inaction} \eea 
where $g^{(6)}_{MN}$ is the six-dimensional metric,  $M_6$ is the six-dimensional 
Planck mass, 
 $M_4$ is the four-dimensional one, $r_c=M_4/M_6^2$ the cross-over scale between 
  four-dimensional and six-dimensional gravity and $\alpha>0$ is the 
  Gauss-Bonnet coupling constant. The singular terms have been written in the particular 
   coordinate system where the metric reads 
\be 
ds^2=g_{\m\n}(x,r)dx^\m dx^\n + dr^2 + L^2(x,r) d\th^2~, 
\label{metric} 
\ee 
where $g_{\mu\nu}(x,0)$ is the brane-world metric 
and $x^{\mu}$ denote four non-compact dimensions, $\mu=0,...,3$, 
whereas $r,\th$ denote the radial and angular coordinates of the 
two extra dimensions (the r direction may 
 or may not  be compact and the $\th$ coordinate ranges form $0$ to $2\pi$). 
Capital $M$,$N$ indices will take values in the six-dimensional 
space. Note, that we have assumed that there exists an azimuthal 
symmetry in the system, so that both the induced four-dimensional 
metric and the function $L$ do not depend on $\th$. 
 The normalization of the $\d$-function is the one discussed in \cite{Leblond:2001xr}.

The full equations of motion that are derived from the above 
action are 
 
\be 
 G^{(6)N}_M + r_c^2 G^{(4)\n}_\m \d^\m_M \d_\n^N {\d(r) \over 2 \pi 
L}  -\a H_M^N={1 \over M_6^4} \left[T^{(B)N}_M+T^{(br)\n}_\m 
\d_M^\m \d^N_\n {\d(r) \over 2 \pi L}\right], 
\label{einsteineq}\ee 
 with $G_M^{(6)~N}=R^{(6)~N}_M-{1 \over 
2}R^{(6)} \d_M^N$ the six-dimensional Einstein tensor, 
$G_\m^{(4)~\n}=R^{(4)~\n}_\m-{1 \over 2}R^{(4)} \d_\m^\n$ the 
four-dimensional Einstein tensor and 
 \bea H_M^N={1 \over 2}\d_M^N 
{\cal L}_{GB}-2R^{(6)}R_M^{(6)N}+4R^{(6)}_{MP}R^{NP}_{(6)} 
+4R^{(6)~~~N}_{KMP}R_{(6)}^{KP} -2R^{(6)}_{MK\Lambda 
P}R_{(6)}^{NK\Lambda P}~, \eea 
 where the Gauss-Bonnet combination is 
\be {\cal L}_{GB}=R^{~2}_{(6)}-4 
R_{MN}^{(6)~2}+R_{MNK\Lambda}^{(6)~2}~. \ee 
 
In order that there are no curvature singularities more severe 
than conical, 
 we will impose certain conditions on the value of the extrinsic curvature $K_{\m\n}=g'_{\m\n}$ on the 
 brane, where the prime denotes derivative with respect to $r$, 
  and on the expansion coefficients of the  function $L$ 
\be L=\b_1(x) r +\b_2(x) r^2 + \b_3(x) r^3 +\dots \ee 
 These 
conditions read \cite{Bostock:2003cv} 
\bea 
K_{\m\n}|_{r=0}&=&0~,\label{condition1}\\ 
\b_1={\rm const.}~~~~&{\rm and}&~~~~\b_2=0~. \eea

Imposing these conditions and keeping only the finite part in 
$L''/L$, the Einstein equations (\ref{einsteineq}) can be 
evaluated at $r=0$. The effective Einstein equations on the brane (obtained by equating the $\d$-function parts of the Einstein equations) 
are \cite{Bostock:2003cv,Janpaper} \be G^{(4)\n}_\m={1 \over 
M_{Pl}^2 }\left[T^{(br)\n}_\m -\Lambda_4\d_\m^\n \right]~, 
\label{4deins}\ee 
with $M_{Pl}^2=M_6^4 [r_c^2+8\pi (1-\b_{1})\a]$ 
and $\Lambda_{4}=-2\pi M_6^4 (1-\b_{1})$. The various components of the bulk Einstein 
equations evaluated at $r=0$ are given in the following:

The $(\m\n)$ component \bea &&G^{(4)\n}_\m -{1 \over 2}K_\m^{\n ~ 
'} +{1 \over 2}\d_\m^\n \left(K'+2{L'' \over L} \right)\nonumber 
\\ &&-\a \left[ {1 \over 2}\d_\m^\n \left(R^{2}-4 
R_{\k\la}^{2}+R_{\k\la\xi\s}^{2}+4K^{\k~'}_\la 
R_\k^\la-2 K' R  -4{L'' \over L}R \right) \right. \nonumber \\ 
&&~~~~~~~~~-2RR_\m^\n+4 R_{\m \k}R^{\n \k}+4R^{~~~~\n}_{\k \m 
\lambda}R^{\k\lambda}-2R_{\m \k\lambda \rho}R^{\n \k\lambda \rho} 
\nonumber \\ &&~~~~~~~~~\left.+4{L'' \over L}R^\n_\m 
+2K'R^\n_\m+K_\m^{\n~'} R-2K_\k^{\n~'} 
R^\k_\m-2g_{\m\la}K^{\la~'}_\k R^{\n\k}\right]= {1 \over 
M_6^4}T^{(B)\n}_\m~. \label{mncomp} \eea 
 
 The $(rr)$ component 
\be 
 -{1 \over 2}R -\a {1 \over 2}(R^{2}-4 
R_{\m\n}^{2}+R_{\m\n\k\lambda}^{2})= {1 \over M_6^4} T^{(B)r}_r~. 
\label{rrcomp} \ee 
 
The $(\th \th) -(rr)$ component \be {1 \over 2}K' +\a \left( K'R - 
2K_\n^{\m~'}R_\m^\n\right)={1 \over M_6^4} (T^{(B)\th}_\th - 
T^{(B)r}_r)~. \label{ththcomp}\ee 
 
The ($\m,r$) component \be T^{(B)r}_\m =0 ~.\ee

In the following, we will study the above equations in a time 
dependent background, assuming that the bulk consists of a pure 
cosmological constant $T^{(B)N}_M=- \La_B \d_M^N$ and that the 
matter content of the brane is an isotropic fluid with 
$T^{(br)\n}_\m=\d_\m^\n ~{\rm diag}(-\r_b,P_b,P_b,P_b)$.

\section{The Constrained Isotropic Case}

We are interested in the cosmological evolution of a flat 
isotropic brane-universe, therefore we will consider the following 
time dependent form of the metric (\ref{metric}) 
\be ds^2= 
-N^2(t,r)dt^2 +A^2(t,r)d\vec x^2 + dr^2 + L^2(x,r) d\th^2~. 
\label{timmetric}\ee 
 We can use the gauge freedom to fix 
$N(t,0)=1$, while we define $A(t,0)\equiv a(t)$. The curvature 
singularity avoidance condition (\ref{condition1}) we imposed, 
dictates that $N'(t,0)=A'(t,0)=0$, while the second derivatives of these metric functions are 
unconstrained. 
 
For this ansatz, the Einstein equations (\ref{4deins}) and 
(\ref{rrcomp}) give

\bea 3{\dot{a}^2 \over a^2}&=&{\rho_{b}+\La_4 \over 
M_{Pl}^2}~,\label{1fried} 
\\ 
2{\ddot{a} \over a } + {\dot{a}^2 \over a^2}  &=&{-P_{b}+\La_4 
\over M_{Pl}^2 }~,\label{2fried} \\ 
 3  \left( {\ddot{a} \over a}+{\dot{a}^2 \over a^2} \right) &+&12 
\a {\ddot{a}\dot{a}^2 \over a^3}= {\La_B \over 
M_6^4}~.\label{friedconstraint} \eea

The equations (\ref{1fried}) and (\ref{2fried}) are the usual 
Friedmann and Raychaudhuri equations of a four-dimensional 
universe with a scale factor $a$, while the third equation 
(\ref{friedconstraint}) appears because of the presence of the 
bulk and acts as a constraint between the matter density and 
pressure on the brane. To see this, a simple manipulation of the 
above equations gives 
 \be 
-{\La_B \over M_6^4}=\left({1 \over 2}+{2 \over 3}\a{\La_4 \over 
M_{Pl}^2}\right) {3P_b-\r_b \over M_{Pl}^2} -2{\La_4 \over 
M_{Pl}^2}\left( 1+ {2 \over 3}\a{\La_4 \over M_{Pl}^2} \right) + 
{2 \over 3}\a {\r_b (3P_b+\r_b ) \over M_{Pl}^2} \label{Priso}~, 
\ee 
which shows a precise relation between $\La_B$, $\r_b$ and 
$P_b$. To simplify the equations, we assume that the vacuum energy (tension) of the brane cancels the  contribution $\La_4$ induced by the deficit angle, \ie 
 \be \r_b=-\La_4 + 
\r_m~~~{\rm and} ~~~ P_b=\La_4  + P_m~, \ee 
 with $P_m=w_c\r_m$. Then 
the above equations read 
 \bea 
&&3{\dot{a}^2 \over a^2}={\rho_m \over M_{Pl}^2}~, \label{1rstfr}\\ 
&&2{\ddot{a} \over a } + {\dot{a}^2 \over a^2}  =-w_c{\r_m \over 
M_{Pl}^2 }~,\label{2ndfr} \eea while the constraint equation 
becomes 
 \be 
-{\La_B \over M_6^4}={\r_m \over M_{Pl}^2}\left[{1 \over 
2}(3w_c-1)+{2 \over 3}(3w_c+1)\a{\r_m \over M_{Pl}^2} 
   \right]~. 
\label{isocon} \ee 
 
Using the metric (\ref{timmetric}), the Einstein equations 
(\ref{mncomp}) and (\ref{ththcomp}) give a set of bulk equations 
which involve the boundary values of the second $r$-derivatives of 
the metric at $r=0$

\bea 
 &&{L'' \over L} \left[ 1 + 12 \a {\dot{a}^2 \over a^2} \right] 
+ 3 {A'' \over A}\left[ 1+4 \a \left(- {\ddot{a} \over 
a}+{\dot{a}^2 \over a^2} \right)\right]=3{\dot{a}^2 \over 
a^2}-{\La_B \over M_6^4} \label{00bulk}~, \\ 
~~\nonumber\\ &&{L'' \over L} \left[ 1 + 4 \a \left(2{\ddot{a} 
\over a}+{\dot{a}^2 \over a^2}\right) \right]+ {N'' \over N} 
\left[ 1 + 4 \a \left(-{\ddot{a} \over 
a}+{\dot{a}^2 \over a^2}\right) \right] \nonumber \\ 
&& ~~~~~~~~~~~~~~~~~~~~~~~~~~~~~~~+ 2 {A'' \over A}\left[ 1+4 \a \left({\ddot{a} \over 
a}-{\dot{a}^2 \over a^2} \right)\right]={\dot{a}^2 \over 
a^2}+2{\ddot{a} \over a}-{\La_B \over M_6^4}~, \label{ijbulk} \\ 
~~\nonumber\\ &&{N'' \over N} \left[ 1 + 12 \a {\dot{a}^2 \over 
a^2} \right] + 3 {A'' \over A}\left[ 1+4 \a \left( 2{\ddot{a} 
\over a}+{\dot{a}^2 \over a^2} \right)\right]=0 \label{rr-thth}~. 
\eea 
 
These equations can be solved for the second $r$-derivatives of 
the metric, as functions of the matter content on the brane, and 
in principle they can give us information about the structure of 
the bulk at $r=0$ 
\bea 
 {A'' \over A}&=&{ {1 \over 4}\left(1+4\a 
{\La_B \over M_{6}^4}\right)(w_c+1){\r_m \over M_{Pl}^2}  \over   1- 2\a 
{\r_m \over M_{Pl}^2}\left[w_c-1+2\a {\r_m \over M_{Pl}^2}(w_c+1)(3w_c-1)   \right]   }~, \label{A''sol}\\ 
{N'' \over N}&=&3 {A'' \over A}~ {4\a w_c{\r_m \over M_{Pl}^2} -1 \over 4 \a {\r_m \over M_{Pl}^2} +1 }~, \label{N''sol}\\ 
{L'' \over L}&=&{1 \over 1+ 4\a {\r_m \over M_{Pl}^2}}\left[ {\r_m 
\over M_{Pl}^2} -{\La_B\over M_{6}^4}  -3 {A'' \over A} 
 \left(1 + 2 \a (w_c+1) {\r_m \over M_{Pl}^2}  \right) 
 \right]\label{L''sol}~. 
\eea

A potential problem in the cosmology of the system would be, if the 
   denominator of \reef{A''sol} is equal to zero, \ie when $w_c=w_s^\pm$ with 
\be w_s^\pm \equiv {-\left(1+4\a{ \r_m \over M_{Pl}^2} \right) \pm 
\sqrt{64 \a^2{ \r_m^2 \over M_{Pl}^4} +32\a{ \r_m \over M_{Pl}^2} 
+13  }  \over 12 \a{ \r_m \over M_{Pl}^2}}~. \label{singular} \ee 
When this happens, the six-dimensional curvature invariant will 
diverge close to the brane. 
 Thus, after discussing the cosmological evolution of the brane world-volume, we should always check that 
  it does not pass through a point which satisfies the above relation.

\section{Cosmological Evolution of an Isotropic  Brane-\\Universe} 
 
From the constraint relation \reef{isocon},  we can solve for 
$w_c$, the allowed equation of state of the  matter on the brane. 
It should satisfy the following equation \be w_c={-2 {\La_B \over 
M_6^4} +{\r_m \over M_{Pl}^2} \left(1- {4 \over 3}\a{\r_m \over 
M_{Pl}^2}\right) \over 3{\r_m \over M_{Pl}^2}\left(1+ {4 \over 
3}\a{\r_m \over M_{Pl}^2}\right) } ~, \label{constrw} \ee with 
$\r_m >0$ so that the Hubble parameter is real.

Before analyzing the system, let us note a first important difference between the system of the pure four dimensional 
  dynamics and the one with the extra constraint added because of the presence of the bulk. 
   In the four dimensional system, a constant $w$ is allowed and its value is preserved during the evolution 
    of the system. On the other hand, the evolution of the system with the extra constraint forbids 
     any evolution with constant $w_c$. Indeed, by differentiating (\ref{constrw}) and using \reef{1rstfr} 
       and the conservation equation $\dot{\r}_m+3(1+w_c)\r_m{\dot{a} \over a}=0$, 
       we can find a differential equation for $w_c$ 
\be \dot{w}_c+3(1+w_c)\r_m {\de w_c \over \de \r_m}{\dot{a} \over 
a}=0~. \label{wdifiso} \ee 
 
Imposing a further condition of keeping $w_c$ constant, would 
result to a constant $\r_m$, related to $w_c$ in a specific way, 
 and by the conservation equation, to zero Hubble for $a$. Thus, an {\it a priori} fixing of $w_c$ 
  would result to an inconsistent system. 
 
To study the cosmological evolution, we look at the system 
\bea 
\dot{H}&=&-{1 \over 2}(1+w_c){\r_m \over M_{Pl}^2}~,\\ 
\dot{\r}_m&=&-3(1+w_c)\r_m H~,\label{conserv1} 
\eea 
 where 
$H=\dot{a}/ a$ is the Hubble parameter. 
 We will analyze the above system of the isotropic case for $\La_B=0$ 
  and $\La_B \neq 0$, because of the different features that arise in the two choices of this parameter.

\subsection{Evolution of the System for $\La_B=0$}

From the above dynamical system, taking into account 
\reef{constrw}, we find that there is only one fixed point in the 
evolution, the one with 
 \be 
(\r_m,H^2,w_c)=(0,0,1/3)~. 
\ee 
Linear perturbation around this point reveals that it is an {\it attractor}. 
 From \reef{wdifiso} and the conservation equation \reef{conserv1} we find 
 for the Hubble parameter 
\be {\dot{a} \over a}={2 \dot{w}_c \over 
(1+3w_c)(1-3w_c)(1+w_c)}~. \ee The above equation can easily be 
integrated and solved for $w_c$ (the $''-''$ sign in the solution 
of $w_c$ is rejected because it gives imaginary  Hubble parameter) 
 
\be 
w_c=-{1 \over 3} + {2 \over 3}~{a^4 \over \sqrt{3+a^8}}~. 
\ee 
 
From this equation, we see that from any initial condition along the line of tuning \reef{constrw}, 
 the expansion of the universe drives the equation of state to $w_c \to 1/3$, \ie radiation. 
  We have also verified this by integrating the system numerically. During this cosmological evolution, 
  it can never happen that $w_c=w_s^\pm$ (compare  \reef{singular} with \reef{constrw}) and thus the whole system is regular.

\subsection{Evolution of the System for $\La_B \neq 0$}

From the dynamical system, as written in the previous subsection, 
taking into account \reef{constrw} with  $\La_B \neq 0$, we find 
that there is a fixed point in the cosmological evolution 
\be 
(\r_m,H^2,w_c)=\left(\r_f,{\r_f \over 3 M_{Pl}^2},-1\right)~, 
\ee 
with ${\a\r_f \over M_{Pl}^2} = -{3 \over 4}+{3 \over 4}\sqrt{1+{4 
\over 3}{\a\La_B \over M_6^4}}$. 
Since we should have $\r_m 
>0$, this fixed point exists only for $\La_B>0$ and 
corresponds to a de Sitter vacuum.  Linear perturbation around 
this point reveals that it is an {\it attractor}. Thus, any matter 
density on the brane eventually evolves to a state of vacuum 
energy.  During this evolution, the singular point $w_c=w_s^-$ can 
be encountered only if $\La_B/M_6^4 \gtrsim .576$ (equating the 
expressions \reef{singular} and \reef{constrw} the energy density 
$\r_m$ can be real and positive only for this range of $\La_B$). 
However, it is easily verified that for these ``dangerous'' values 
of $\La_B$, it is $w_s^- < -1$. Thus, even for those values of 
$\La_B$ for which the system is singular, the dangerous point 
$w_s^-$ is reached only if initially $w_c < w_s^- < -1$,  \ie for 
rather unphysical initial equations of state. All other evolutions 
will flow to the fixed point without ever passing though $w_s^-$. 
 
\begin{figure}[t!] 
 
\begin{center} 
\epsfig{file=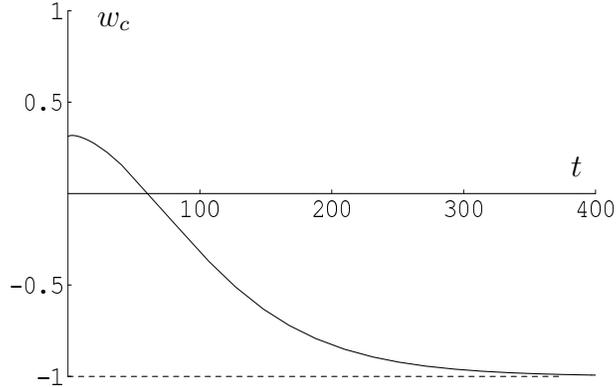,width=8cm,height=5cm} 
 
\begin{picture}(.01,.01)(0,0) 
 
\Text(100,95)[c]{$t$}

\Text(-75,150)[c]{$w_c$}

\end{picture} 
 
\vskip-5mm 
 
\caption{The  evolution of the equation of state $w_c$ for the case of $\a \La_B / M_6^4 = 10^{-4}$ with initial conditions $w_c=.31$ and ${\a \r_m \over M_{Pl}^2} =.02$.} 
 
\label{smallLBiso} 
\end{center} 
\end{figure}

  A potentially interesting case, with a cosmological evolution 
resembling that of our universe, would be the one where 
\mbox{$0<\a\La_B / M_6^4 \ll 1$}. Then, the line of isotropic 
tuning, as it is given by relation \reef{constrw}, has a maximum 
close to $w_c \sim 1/3$. The evolution of an initial energy 
density  larger than the one corresponding to that  maximum, will 
evolve towards $w_c \sim 1/3$, pass from  $w_c=0$ and asymptote to 
$w_c \to -1$ (see Fig.~\ref{smallLBiso} for an example). The 
asymptotic  value of the  effective cosmological constant  at the 
fixed point will be $\a {\La_{eff} \over M_{Pl}^2}  \sim \a {\La_B 
\over M_6^4}\ll 1$ and should be rather small to match with 
observations (when performing such a comparison,  it is reasonable 
to assume  that all the dimensionful scales of the theory are 
roughly of the same order, \ie  $\a^{-1/2} \sim M_{Pl} \sim M_6$). 
The latter requirement of extremely small $\a\La_B / M_6^4$ is the 
usual cosmological constant problem.  Although the standard 
cosmological evolution is described by piecewise constant 
equations of state with $w \sim 1/3,0,-1$, in the present theory 
the  equation of state has always time dependence.

For $\La_B < 0$, there does not exist any fixed point. The 
evolution of this system has a runaway behaviour and  flows to 
$w_c \to \infty$ while $\r_m \to 0^+$.  During this evolution, the 
singular point $w_c=w_s^+$ can be encountered only if  $\La_B/M_6^4 
\lesssim -.326$ (equating the expressions \reef{singular} and 
\reef{constrw} the energy density $\r_m$ can be real and positive 
only for this range of $\La_B$). It is easily verified that for 
these ``dangerous'' values of $\La_B$, it is $w_s^+ > 1/3$. 
Therefore even in the case in which $\La_B$ takes these values, 
the dangerous point $w_s^+$ is reached only if initially  $w_c < 
w_s^+$,

We finally note that, for  $\La_B \neq 0$, there does not exist a 
fixed point with  $(\r_m,H^2)=(0,0)$ because the equation of state 
$w_c$ diverges at that point.

\section{The Unconstrained Anisotropic Case} 
 
In the previous section we found some interesting cosmological 
evolutions of an isotropic brane-universe if $\La_B>0$. We should 
keep in mind however, that in all cases studied, the energy 
density on the brane is tuned to the equation of state 
  in a specific way, which seems at first sight artificial.  Therefore, it would be worth studying some cosmological evolution, in a codimension-2 brane-world model, 
   in which this tuning is not required. If the system then evolves towards the previously studied line of isotropic tuning, we will conclude that this tuning is an attractor and thus not artificial.

To do so, we have to consider geometries which are not isotropic 
 and in which the Riemann tensor cannot be expressed in terms of the Ricci tensor and the curvature scalar \cite{Janpaper}. 
  Then,  the constraint equation \reef{rrcomp} will not give rise to a brane-bulk matter tuning, 
   but rather to a dynamical equation for the anisotropy of the space. 
    For this purpose, let us consider the following anisotropic ansatz where the metric functions depend only on the time $t$ and the radial coordinate $r$ (\ie keeping the azimuthal symmetry) 
\be 
 ds^2= -N^2(t,r)dt^2 +\sum_{i=1}^3 A_i^2(t,r)(d x^i)^2 + dr^2 + 
L^2(x,r) d\th^2~. 
\label{anisometric} 
\ee 
 We can again use the gauge freedom to fix 
$N(t,0)=1$, while we define $A_i(t,0)\equiv a_i(t)$. 
 The singularity conditions dictate as before that $N'(t,0)=A'_i(t,0)=0$, 
  while the second derivatives of these metric functions   are unconstrained. The most general anisotropic evolution scale factors with the above property can be written as 
\bea 
&&A_1(t,r)=a(t)b(t)c(t)+\xi_1(t)r^2 + \dots ~,\\ 
&&A_2(t,r)={a(t) \over b(t)}+\xi_2(t)r^2 + \dots~, \\ 
&&A_3(t,r)={a(t) \over c(t)}+\xi_3(t)r^2 + \dots \eea where 
$a=(a_1a_2a_3)^{1/3}$  represents the ``mean'' scale factor and 
$b$, $c$ represent two degrees of anisotropy.  To simplify further the analysis of the system, we will choose $c=const.$ 
  (by a coordinate redefinition then we can always set $c=1$). 
    The dynamics of this particular choice  can help us to understand the 
   qualitative features of the general case.

Note  that, with the choice $c=const.$, we have $A_1(t,0)A_2(t,0)=A_3^2(t,0)$, 
but not also $A_1(t,r)A_2(t,r)=A_3^2(t,r)$. The later relation would lead to extra 
constraints which will overdetermine the system. 
  For this ansatz, the Einstein equations (\ref{4deins}) and 
(\ref{rrcomp}) give

\bea 
3{\dot{a}^2\over a^2} - {\dot{b}^2 \over b^2}&=&{\r_b +\La_4 \over M_{Pl}^2 }~, \label{00}\\ 
2{\ddot{a} \over a}+{\dot{a}^2 \over a^2}-{\ddot{b} \over b} 
+2{\dot{b}^2 \over b^2} &-&3 {\dot{a}\dot{b} 
 \over a b} ={-P_b +\La_4 \over M_{Pl}^2}~, \label{11}\\ 
2{\ddot{a} \over a}+{\dot{a}^2 \over a^2}+{\ddot{b} \over b} &+&3 
{\dot{a}\dot{b} \over a b}={-P_b 
 +\La_4 \over M_{Pl}^2  }~, \label{22}\\ 
2{\ddot{a} \over a}+{\dot{a}^2 \over a^2}&+&{\dot{b}^2 \over b^2}={-P_b +\La_4 \over M_{Pl}^2  }~, \label{33}\\ 
3{\ddot{a} \over a}+3{\dot{a}^2 \over a^2}+ {\dot{b}^2 \over 
b^2}&+&4\a \left[ 3 {\ddot{a}\dot{a}^2 \over a^3} 
 + 2 {\dot{a}\dot{b}^3 \over a b^3} -2 {\dot{a}^2\dot{b}^2 \over a^2 b^2} -2{\ddot{b}\dot{b}\dot{a} 
  \over b^2a}- {\ddot{a}\dot{b}^2 \over a b^2} \right] ={\La_B \over M_6^4 }~. \label{rrcon} 
\eea

The equations \reef{00}, \reef{11}, \reef{22}, \reef{33} are the 
equations of  four dimensional Einstein gravity with the 
previously postulated anisotropy, while the last equation 
(\ref{rrcon}) appears because of the presence of the extra 
dimensions and corresponds to the constraint equation of the 
isotropic case that we considered. It is easy to see that from the 
three equations \reef{11}, \reef{22}, \reef{33}, one is redundant. This happens because we have frozen the $c$ degree of freedom.  Keeping then two linear combinations of them, the brane 
Einstein equations take a simpler form \bea 
3{\dot{a}^2\over a^2} - {\dot{b}^2 \over b^2}&=&{\r_b +\La_4 \over M_{Pl}^2}~, \label{hubbleb}\\ 
2{\ddot{a} \over a}+4 {\dot{a}^2 \over a^2}&=&{\r_b-P_b +2\La_4 \over M_{Pl}^2}~, \label{accelb}\\ 
{\ddot{b} \over b}-{\dot{b}^2 \over b^2}+3{\dot{a}\dot{b} \over a b}&=&0~. \label{bb} 
\eea 
 
These are purely four-dimensional equations, while equation 
\reef{rrcon} coming from the extra dimensions is now dynamical, 
providing a Hubble equation for $b$ \be 
\renewcommand{\arraystretch}{1.5} 
{\dot{b}^2 \over b^2}=-{\r_b +\La_4 \over 4 M_{Pl}^2 } \pm {\sqrt{3} \over 16\a}\sqrt{\begin{array}{l}-16\a{\La_4\over M_{Pl}^2} \left(2 + \a {\La_4 \over M_{Pl}^2}\right) +8\a \left[ 2 {\La_B \over M_6^4} + {\r_b \over M_{Pl}^2}\left(-1+2\a {\r_b \over M_{Pl}^2}\right)\right.\\ \left.\phantom{1}~~~~~~~~~~~~~~~~~~~~~~~~~~~~~~~~~~~~~~~~~~+2 {P_b \over M_{Pl}^2}\left({3 \over 2} +2\a {\r_b +\La_4 \over M_{Pl}^2}\right)\right]\end{array}}~. 
\ee

We will now assume, as in the previous sections, that the vacuum energy (tension) 
 of the brane cancels the  contribution $\La_4$ induced my the deficit angle, \ie that 
 
\be \r_b=-\La_4 + \r_m~~~{\rm and} ~~~ P_b=\La_4  + P_m~, \ee with 
$P_m=w\r_m$. After this simplification, the above equations read 
\bea 
3{\dot{a}^2\over a^2} - {\dot{b}^2 \over b^2}&=&{\r_m \over M_{Pl}^2}~, \label{Heq}\\ 
2{\ddot{a} \over a}+4 {\dot{a}^2 \over a^2}&=&(1-w){\r_m \over M_{Pl}^2}~, \label{addeq}\\ 
{\ddot{b} \over b}-{\dot{b}^2 \over b^2}+3{\dot{a}\dot{b} \over a 
b}&=&0~, \label{bddeq} \eea while the equation coming from the 
extra dimensions becomes 
\be 
 {\dot{b}^2 \over b^2}=-{\r_m \over 4 
M_{Pl}^2 } + \sqrt{{3 \over 32\a}}\sqrt{2{\La_B \over M_6^4}+ 
{\r_m \over  M_{Pl}^2 }\left[ (3w-1)+2(2w+1)\a {\r_m \over 
M_{Pl}^2 }   \right]} \equiv f(\r_m,w)~. \label{fdef} 
\ee 
 [The $''-''$ sign in front of the square root has been rejected, since 
it always gives rise to imaginary Hubble for either $a$ or $b$.] 
 It is interesting to observe that the Hubble equation 
\reef{Heq} for the ``mean'' scale factor $a$, after substitution of \reef{fdef}, has apart 
from  the usual linear term in $\rho$ (of the conventional 
four-dimensional cosmology), additional  correction terms in 
$\rho$. This is similar to what happens also to five-dimensional 
brane-world models \cite{5d,pap}  and is  due to the presence of 
extra dimensions. This modification happens only in the 
anisotropic case.  In the pure isotropic case the four-dimensional 
brane-universe feels the extra dimensions by only adjusting its 
energy density to its equation of state,  but without any 
modification in  the structure of the Friedmann equation.

The Einstein equations 
(\ref{mncomp}) and (\ref{ththcomp}) give again a set of bulk equations 
which involve the boundary values of the second $r$-derivatives of 
the metric at $r=0$. It is analogous to the system \reef{00bulk}, 
 \reef{ijbulk}, \reef{rr-thth} and has solution identical with the isotropic case \reef{A''sol}, 
  \reef{N''sol}, \reef{L''sol} with $A''_i/A_i = A''/A$. As noted there, 
   we should make sure that any evolution of the system does not pass through 
    the points with $w=w_s^\pm$, where $w_s^\pm$ given by \reef{singular}, 
     since then the quantities  $A_i''/A_i$, $N''/N$,  $L''/L$ will diverse. 
      In addition, because in the anisotropic case $\r_m<0$ is 
      allowed, as it can be seen from equation \reef{Heq}, 
       evolutions which pass from $\a\r_m /M_{Pl}^2=-1/4$ will give singular 
         $N''/N$ and  $L''/L$, and therefore should be avoided.

\section{Cosmological Evolution of an Anisotropic Brane-Universe}

Before analyzing the system, let us note again that in contrast to  pure four dimensional anisotropic dynamics, 
 where  a constant $w$ is allowed, in the present case, where there is an extra dynamical 
  equation because of the presence of the bulk,  $w$ has to evolve. 
   Indeed, by differentiating (\ref{fdef}) 
    and then using the four dimensional equations of motion 
     and the conservation equation $\dot{\r}_m+3(1+w)\r_m{\dot{a} \over a}=0$ 
we can find a differential equation for $w$ 
\be 
 {\de f \over \de 
w} \dot{w}+3\left[2f-(1+w)\r_m {\de f \over \de 
\r_m}\right]{\dot{a} \over a}=0~. 
\label{wdifaniso} 
\ee 
 Imposing a 
further condition to keep $w$ constant, would result to a constant 
$\r_m$ related to $w$ in a specific way and by the conservation 
equation, to zero Hubble for $a$. Thus, an {\it a priori} fixing 
of $w$ would result to an inconsistent system. 
 
To have real Hubble 
 parameters  for  $a$ and $b$, $\r_m$ and $w$ 
have to lie in specific regions for which the following 
inequalities are satisfied 
\be f>0~~~~~,~~~~~f+{\r_m \over 
M_{Pl}^2}>0~. \label{allowedf} \ee 
 Define the following boundaries 
of the allowed regions \bea w_1&\equiv&{-2 {\La_B \over M_6^4} 
+{\r_m \over M_{Pl}^2} \left(1- {4 \over 3}\a{\r_m \over 
M_{Pl}^2}\right) \over 3{\r_m \over M_{Pl}^2}\left(1+ {4 \over 
3}\a{\r_m \over M_{Pl}^2}\right) }~, 
\label{w1}\\ 
w_2&\equiv&{-2 {\La_B \over M_6^4} +{\r_m \over M_{Pl}^2} 
\left(1+4 \a{\r_m \over M_{Pl}^2}\right) \over 3{\r_m \over 
M_{Pl}^2}\left(1+ {4 \over 3}\a{\r_m \over M_{Pl}^2}\right) }~. 
\label{w2} \eea The quantity $w_1$ coincides with the line of 
isotropic tuning $w_c$ given in \reef{constrw}. The inequalities 
\reef{allowedf} can then be re-expressed as conditions for $\r_m$ 
and $w$: 
 
\begin{itemize} 
 
\item For $\r_m>0$, we should have $w>w_1$. 
 
\item For $\r_m<0$, we distinguish two cases 
 
\subitem - for $-{3 \over 4}<{\a \r_m \over M_{Pl}^2}<0$, we should have $w<w_2$. 
 
\subitem - for ${\a \r_m \over M_{Pl}^2}<-{3 \over 4}$, we should have $w>w_2$.

\end{itemize}

To study the cosmological evolution, we look at the system 
\bea 
\dot{H_a}&=&-{1 \over 2}(1+w){\r_m \over M_{Pl}^2}-f~,\label{Ha}\\ 
\dot{H_b}&=&-3H_a H_b~,\label{Hb}\\ 
\dot{\r}_m&=&-3(1+w)\r_m H_a~, \label{conserv} 
\eea 
 where $H_a=\dot{a}/ 
 a$ and $H_b=\dot{b} / b$ are  the Hubble parameters 
for $a$ and $b$ respectively. The third equation is the energy 
conservation equation in which only the Hubble parameter $H_{a}$ 
for the ``mean''  scale factor $a$ appears.  Again we will analyze 
the anisotropic case for $\La_B=0$ and $\La_B \neq 0$ and we will 
compare the cosmological evolution with the cosmological evolution 
of the tuned isotropic case.

\begin{figure}[t]

\epsfig{file=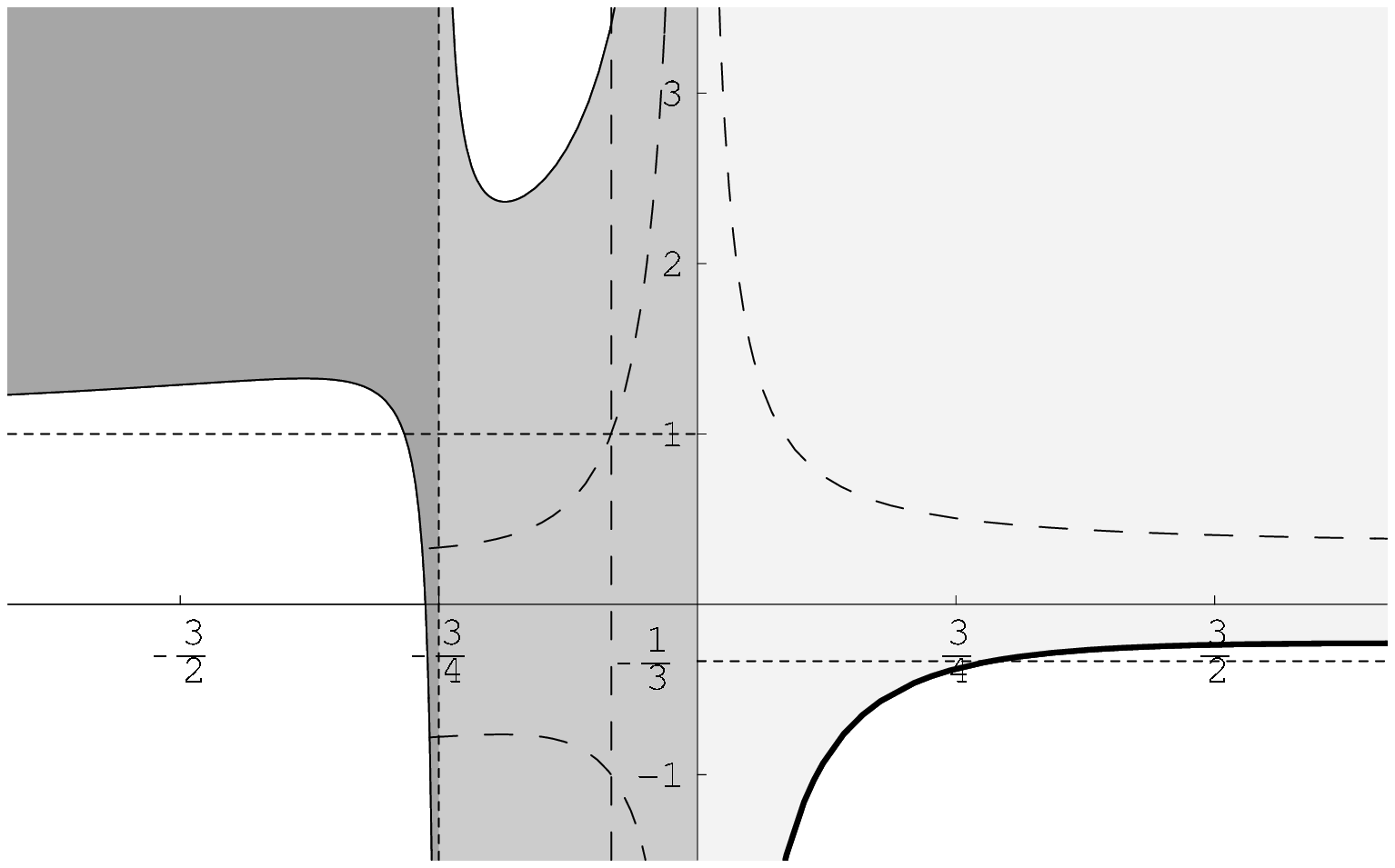,width=8cm,height=5cm} 
\epsfig{file=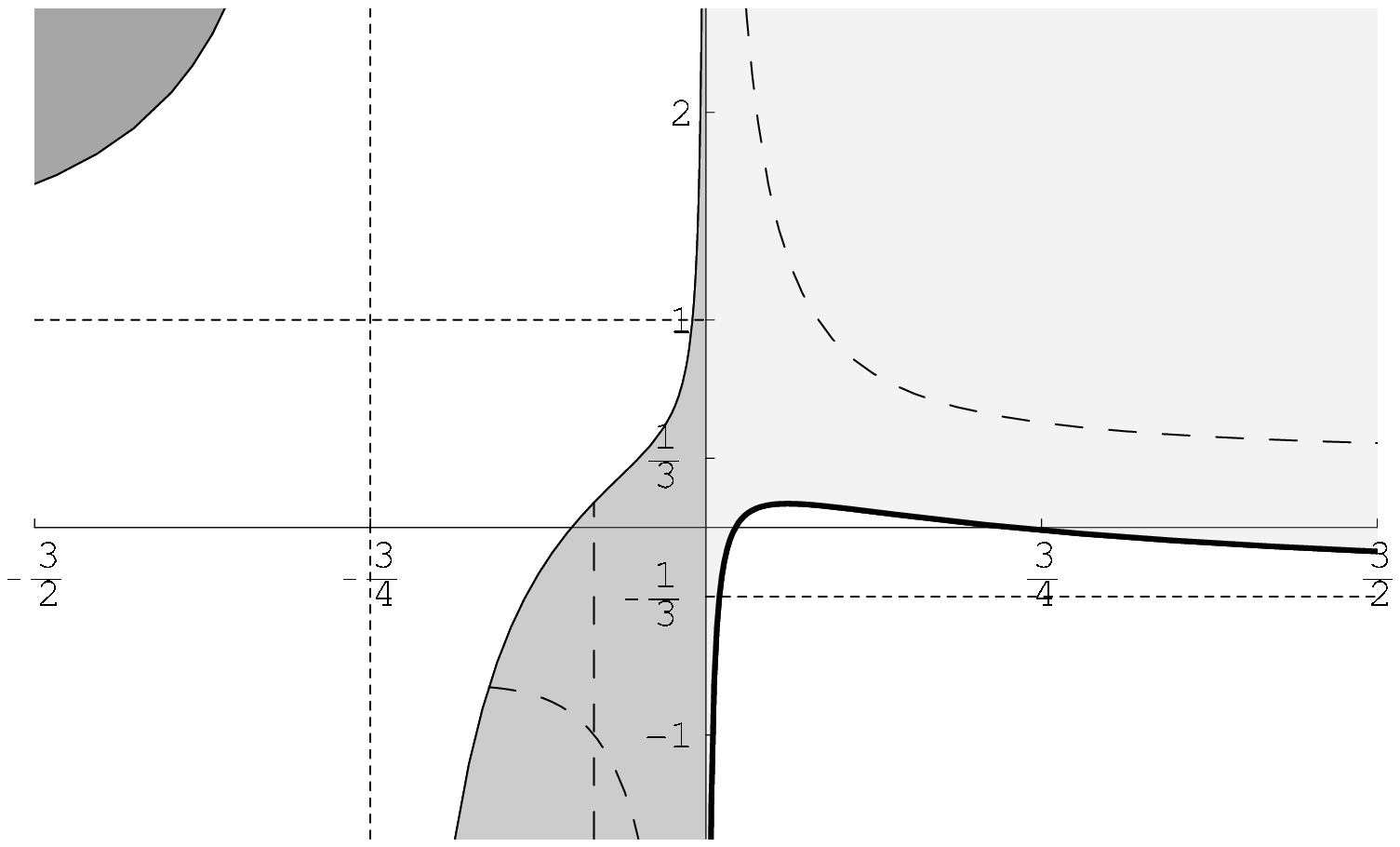,width=8cm,height=5cm} \\ 
\epsfig{file=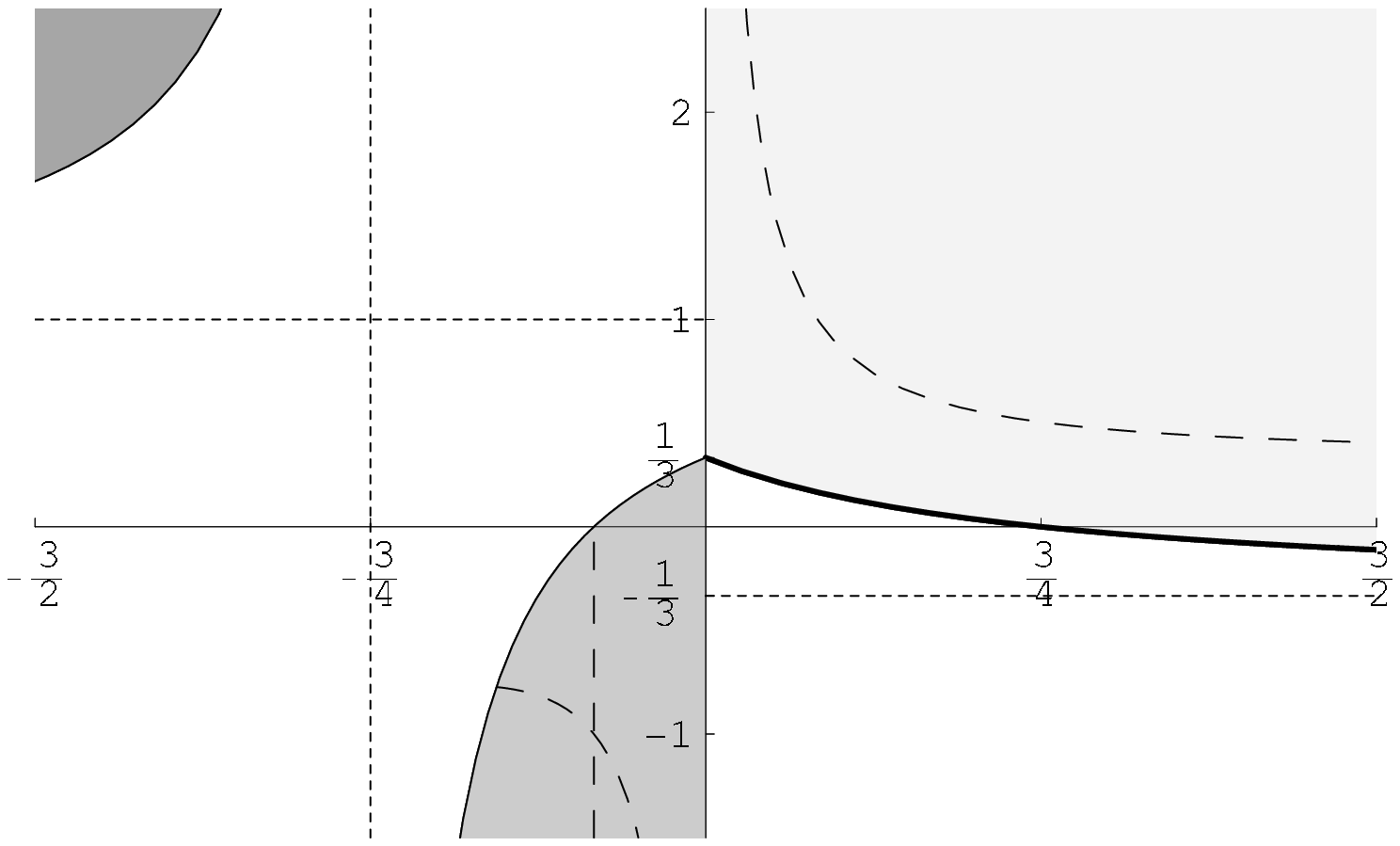,width=8cm,height=5cm} 
\epsfig{file=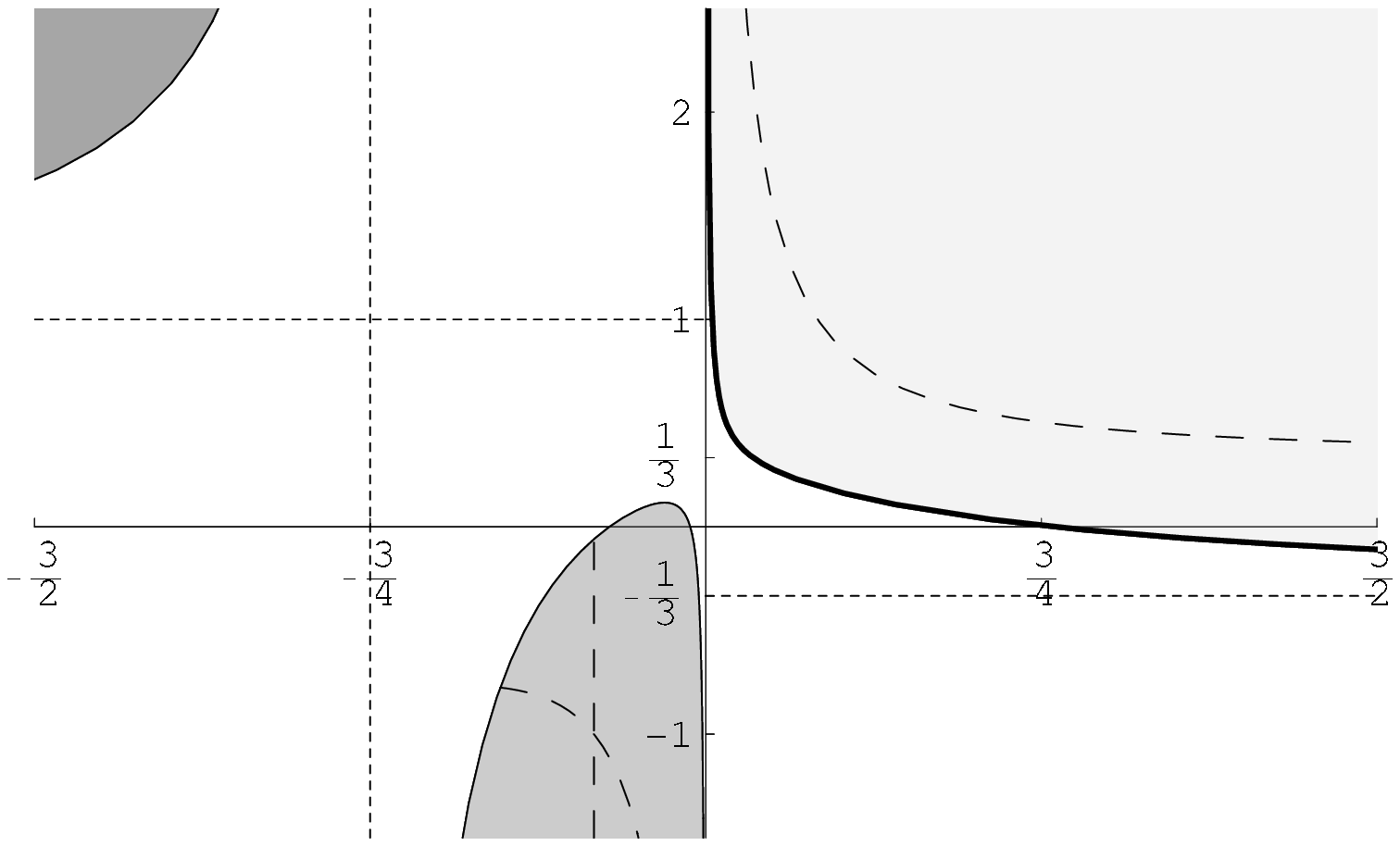,width=8cm,height=5cm} 
 
 \begin{center} 
\begin{picture}(.01,.01)(0,0) 
 
\Text(215,93.5)[c]{${\a \r_m \over M_{Pl}^2}$} 
\Text(219,236.7)[c]{${\a \r_m \over M_{Pl}^2}$} 
\Text(-13,222.3)[c]{${\a \r_m \over M_{Pl}^2}$} 
\Text(-15,93.5)[c]{${\a \r_m \over M_{Pl}^2}$} 
\Text(-108,305)[c]{$w$} 
\Text(128,305)[c]{$w$} 
\Text(128,160)[c]{$w$} 
\Text(-105,160)[c]{$w$} 
 
\Text(170,275)[c]{{\bf I}} 
\Text(-60,275)[c]{{\bf I}} 
\Text(-60,130)[c]{{\bf I}} 
\Text(170,130)[c]{{\bf I}}

\Text(-140,250)[c]{{\bf II}} 
\Text(102,200)[c]{{\bf II}} 
\Text(104,55)[c]{{\bf II}} 
\Text(-130,55)[c]{{\bf II}}

\Text(-217,160)[c]{{\bf III}} 
\Text(17,160)[c]{{\bf III}} 
\Text(13,302)[c]{{\bf III}} 
\Text(-200,280)[c]{{\bf III}} 
 
 \Text(250,110)[c]{(d)} 
\Text(250,250)[c]{(b)} 
\Text(-250,110)[c]{(c)} 
\Text(-250,250)[c]{(a)}

\Vertex(-97,184){3} 
\Vertex(115,188){3} 
 \Vertex(-117,92){3} 
 
\end{picture} 
 
\vskip-5mm 
 
\caption{The three regions of allowed $\r_m$ and $w$ for:  (a) ${\a\La_B \over M_6^4}>{3 \over 4}$,  (b)   $0<{\a\La_B \over M_6^4}<{3 \over 4}$, (c) ${\a\La_B\over M_6^4}=0$ and (d)   ${\a\La_B \over M_6^4}<0$. The thick lines  correspond to the tunings  $w=w_c$ of the isotropic case. The thick dots represent the fixed point attractors of the evolution. The lines of long dashing denote the singular points $w=w_s^\pm$ and $\a \r_m / M_{Pl}^2=-1/4$.} 
\label{regionsL} 
\end{center} 
\end{figure}

\subsection{Evolution of the System with $\La_B=0$}

There are three different regions in the $(w,\r_m)$ plane where these inequalities are satisfied 
 (see  Fig.~\ref{regionsL}c).  This relative freedom to choose the matter on the brane should 
  be compared to the tuning that happens in the isotropic case. 
  Relaxing the isotropy condition, the system can have initial 
   conditions in a vast region of the parametric space. 
    The line of isotropic tuning is just  the boundary 
     of region I. Furthermore, in the anisotropic case we have 
      two other allowed regions for $\r_m<0$ where evolution is also possible.

From the dynamical system \reef{Ha}, \reef{Hb}, \reef{conserv} we see that there is only one fixed point and that it is the same with that of the isotropic evolution, \ie 
 \be 
(\r_m,H^2_a,H^2_b,w)=(0,0,0,1/3)~. 
\ee 
Linear perturbation around this point reveals again that it is an {\it attractor}. 
 
The presence of the previous attractor fixed point will drive the 
system towards a final isotropic  state of radiation. The way in 
which this fixed point is approached from an arbitrary initial 
energy density, can tell us whether the line of isotropic tuning 
is an attractor or not. If the anisotropy monotonically decreases 
during the evolution, it means the the line of isotropic tuning is 
an attractor.

In order to analyze the features of the anisotropic evolutions, we proceed numerically. We solve the system of the two second order equations 
\reef{addeq}, \reef{bddeq} and the two first order equations \reef{wdifaniso}, 
\reef{conserv} for the four functions $a$, $b$, 
$\r_m$, $w$.  The initial conditions for $\r_m$ and $w$ are such 
that they lie in the allowed regions of Fig.~\ref{regionsL}c and 
the initial conditions for 
 $\dot{b}$, $\dot{a}$ are such that \reef{Heq} and \reef{fdef} are satisfied. We can check that the later two conditions are satisfied throughout the numerical evolution of the system, although imposed only once initially. 
 
To understand how the anisotropy involves we define  the mean 
anisotropy  by the following quantity 
\be 
 A=\sqrt{\sum_{i=1}^3 
{(\langle H \rangle -H_i)^2 \over 3 \langle H\rangle^2}}=\sqrt{{2 
\over 3}}\left|{a \dot{b} \over \dot{a}b}\right|~, 
\ee 
 where 
$H_i=\dot{a}_i/ a_i$ (with $a_i$ defined after \reef{anisometric}) 
and $\langle H\rangle={1 \over 3}\sum_{i=1}^3 H_i=\dot{a}/ a$~. 
 
\begin{figure}[t]

\begin{center} 
\epsfig{file=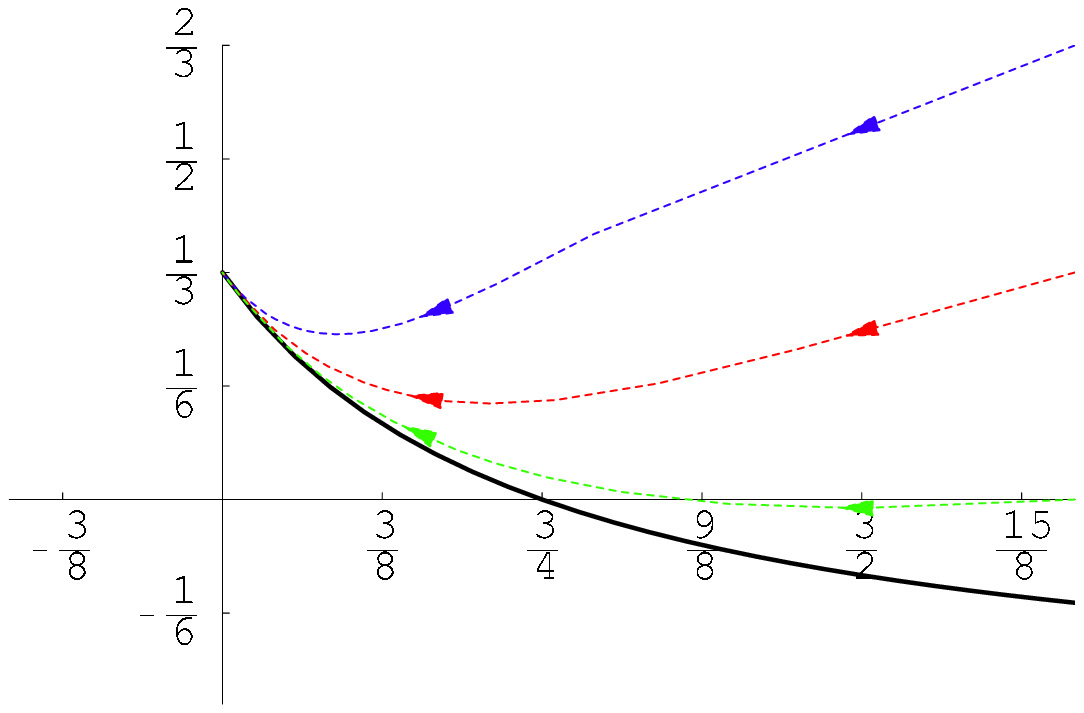,width=8cm,height=5cm} 
\epsfig{file=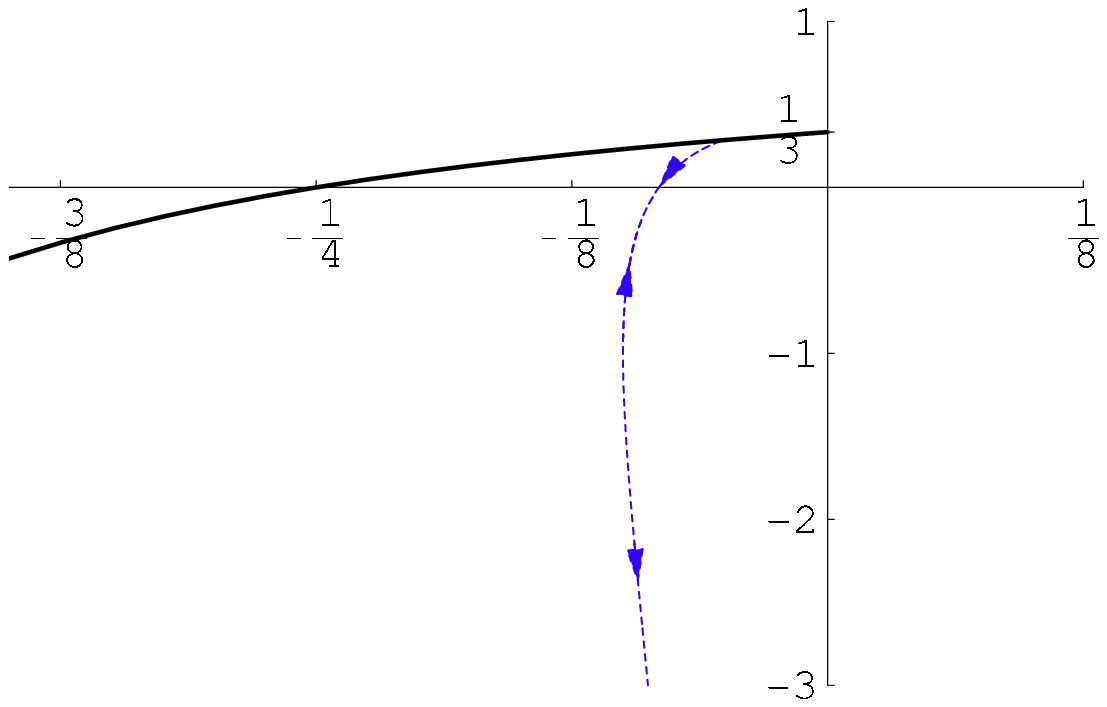,width=8cm,height=5cm} 
\epsfig{file=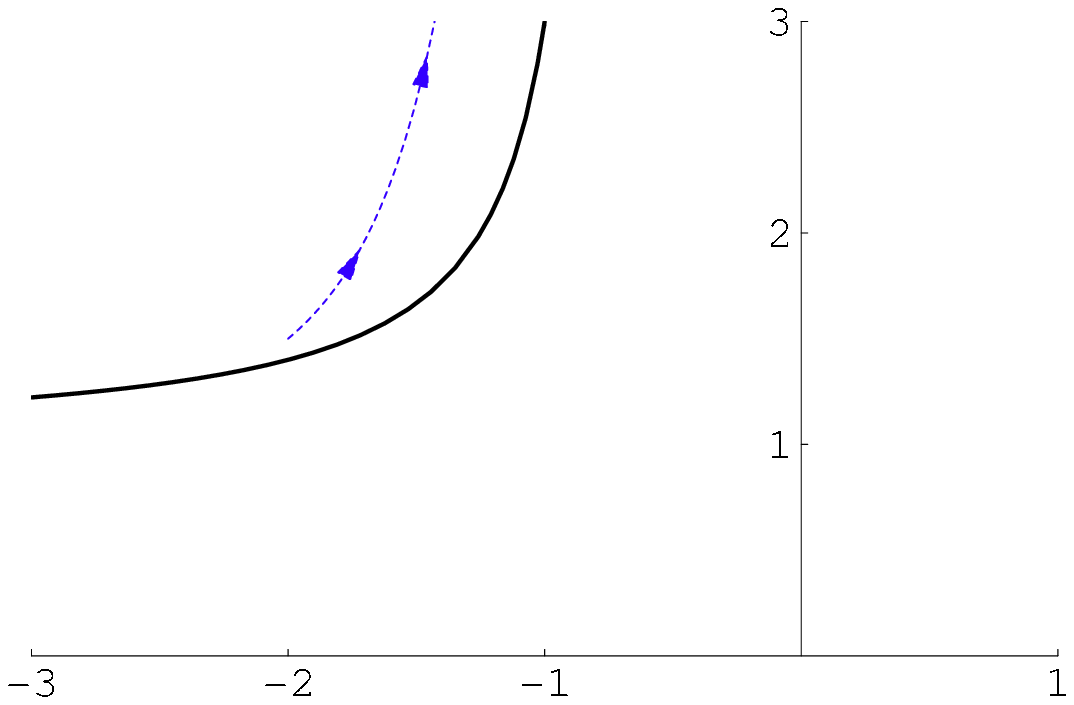,width=8cm,height=5cm} 
 
\begin{picture}(.01,.01)(0,0) 
 
\Text(105,35)[c]{${\a \r_m \over M_{Pl}^2}$} 
\Text(215,275)[c]{${\a \r_m \over M_{Pl}^2}$} 
\Text(-10,215)[c]{${\a \r_m \over M_{Pl}^2}$}

\Text(70,145)[c]{$w$} 
\Text(185,290)[c]{$w$} 
\Text(-170,285)[c]{$w$}

\Vertex(-183,245){3}

\Text(-5,199)[c]{\Green{$\bigstar$}} 
\Text(-5,245)[c]{\Red{$\bigstar$}} 
\Text(-5,291)[c]{\Blue{$\bigstar$}} 
\Text(130,229)[c]{\Blue{$\bigstar$}} 
\Text(-53,90)[c]{\Blue{$\bigstar$}}

\Text(-70,120)[c]{{\bf III}} 
\Text(90,220)[c]{{\bf II}} 
\Text(-90,240)[c]{{\bf I}}

\end{picture} 
 
\vskip-5mm 
 
\caption{The time evolution for $\La_B=0$ in the three regions for different initial conditions: $w_0=0,1/3,2/3$ for the region I, $w_0=-1$  for the region II and $w_0=3/2$ for region III. The thick lines correspond to the boundaries \reef{w1}, \reef{w2} of the three regions and the stars to the initial conditions. The thick dot represents the fixed point of the  evolution. 
 \label{evol}} 
\end{center} 
 
\end{figure}

 Our analysis shows that in region I, whatever the initial 
conditions are, the system slowly tracks the line of isotropic 
tuning and 
 eventually evolves towards $w \to 1/3$ (see 
case I in Fig.~\ref{evol}). The anisotropy decreases during the 
evolution and thus  the tuning between $\r_m$ and $w$ in the 
isotropic case is  an attractor. However, it is a {\it rather weak 
attractor}. The reason for this is that, as we find numerically, 
the anisotropy falls like $A \sim t^{-1/2}$, while at the same 
time the energy density is redshifting  much faster $\r_m \sim 
t^{-2}$. This is also evident from the example of the case I in 
Fig.~\ref{evol}, where the tendency to approach the line of 
isotropic tuning is very weak. Asymptotically,  when the fixed 
point is approached, we have  $A \to 0$, as expected.  These 
cosmological evolutions give rise to a regular geometry  at $r=0$ 
for initial conditions $w<w_s^+$ (see Fig.~\ref{regionsL}c).

The time evolutions for regions II (see case II in Fig.~\ref{evol}) 
and III (see case III in Fig.~\ref{evol}) have on the other hand 
different characteristics. This is because there is no 
corresponding isotropic fixed points towards they could evolve.

In region II, the system evolves towards the boundary line, is 
reflected back and evolves asymptotically to $\r_m \to 0^-$ at $w 
\to - \infty$. The anisotropy increases as the boundary line is 
approached and decreases $A \to \sqrt{2}$ at the asymptotics. Let 
us note also that although region II is connected to the fixed point $w=1/3$, we 
have seen no evolution towards  this point. These cosmological evolutions give rise to a regular geometry  at $r=0$ for initial conditions  on the left of the lines of long dashing in the region II of Fig.~\ref{regionsL}c. 
 
In region III, the system evolves towards the boundary line 
as $w \to \infty$ and $\a \r_m / M_{Pl}^2 \to -3 /  4$. The 
anisotropy decreases  but tends to a constant value $A > 
\sqrt{2}$.  These cosmological evolutions give  always rise to a regular geometry  at $r=0$. 
 
From the above analysis we conclude that, the relaxation of the 
tuning relation between $w$ and $\r_m$, has as a consequence the 
appearance of new branches of brane world evolution (regions II, 
III), while in region I the system tends quickly to the isotropic fixed point 
attractor.  Furthermore, we observe that the anisotropy in all 
cases decreases much more quickly than in the four dimensional 
case with the same initial conditions but without the extra 
dimensional constraint.

\subsection{Evolution of the System with $\La_B \neq 0$} 
 
Studying the asymptotics for $w_1$ and $w_2$ from \reef{w1}, 
\reef{w2} we find that for $\La_B \neq 0$, there are three 
intervals of $\La_B$ with different shape of the allowed regions in the $(\r_m,w)$ plane. The parameter space for these distinct cases is shown in Figs.~\ref{regionsL}a,b,d.

From the dynamical system \reef{Ha}, \reef{Hb}, \reef{conserv} we see that there are two fixed points in general. The first one is the same with that of the isotropic evolution, \ie 
 \be 
(\r_m,H^2_a,H^2_b,w)=\left(\r_f,{\r_f \over 3 M_{Pl}^2},0,-1\right)~, 
\ee 
 with ${\a\r_f \over M_{Pl}^2} = -{3 \over 4}+{3 \over 4}\sqrt{1+{4 
\over 3}{\a\La_B \over M_6^4}}$ and it exists only for $\La_B>0$.  Linear perturbation around this point reveals again that it is an {\it attractor}. 
 
The second fixed point that we find is 
 \be 
(\r_m,H^2_a,H^2_b,w)=\left(-\La_B{M_{Pl}^2 \over M_6^4},0,{\La_B \over M_6^4},1\right)~. 
\ee 
and it exists only for $\La_B>0$.  Linear perturbation around this point reveals that it is a {\it repeler}.

 Whenever  the previous attractor fixed point exists, the system 
will be  driven towards a final isotropic  de Sitter state. The 
way in which this fixed point is approached from an arbitrary 
initial energy density, can tell us whether the line of isotropic 
tuning is an attractor or not.  To analyze the features of the 
anisotropic evolutions, we will again proceed numerically.

For ${\a\La_B \over M_6^4}>{3 \over 4}$, the three allowed regions 
of the $(\r_m,w)$ plane are connected (see Fig.~\ref{regionsL}a). 
  However, we can see numerically that any evolutions with 
initial conditions in one of the three regions, stays in that 
region. In other words, there can be no crossing of the ${\a \r_m 
\over M_{Pl}^2}=-{3 \over 4}$ or ${\a \r_m \over M_{Pl}^2}=0$ 
lines.  In region I the system  evolves quickly towards the line 
of isotropic tuning  and tracks it until the attractor fixed point 
of $w=-1$ is reached. We can see numerically that anisotropy falls 
to zero like $A \sim \exp\left[-{\cal O}(1) \sqrt{{\a \La_B \over 
M_6^4}}~t\right]$ and thus the line of isotropic tuning is {\it 
attracting very strongly} the evolution of the system towards it. 
 In region II, the system 
evolves towards $\r_m \to 0^-$ and $w \to + \infty$. Its 
anisotropy decreases and in the asymptotics it tends to $A \to 
\sqrt{2}$.  In region III the system evolves to $w \to + \infty$ 
and  $\a \r_m / M_{Pl}^2 \to -3 /  4$.  The anisotropy decreases, 
but tends to a constant value $A 
> \sqrt{2}$.

 For $0<{\a\La_B \over M_6^4}<{3 \over 4}$, 
 two of the  allowed regions of the $(\r_m,w)$ plane are connected (see Fig.~\ref{regionsL}b). 
  We can again verify numerically that any evolutions with initial conditions in one of these two regions, 
   stays in that region, so that there is no crossing of the ${\a \r_m \over M_{Pl}^2}=0$  line. 
    In region I the system tracks the line of isotropic tuning  and evolves  towards the attractor fixed point 
     of $w=-1$ while it isotropises ($A \to 0$). The anisotropy falls to zero again 
      like  $A \sim \exp\left[-{\cal O}(1) \sqrt{{\a \La_B \over M_6^4}}~t\right]$ and  thus, 
       the line of  isotropic tuning is an {\it attractor}. However, the strength of the attractor 
        can be very weak in the cases when  ${\a\La_B \over M_6^4} \ll1 $. In region II, 
        the system evolves towards $\r_m \to 0^-$ and $w \to + \infty$. Its anisotropy 
        decreases and in the asymptotics it tends to $A \to \sqrt{2}$. 
        In region III the system again evolves to $w \to + \infty$ and 
          $\a \r_m / M_{Pl}^2 \to -3 /  4$.  The anisotropy decreases, 
          but tends to a constant value $A > \sqrt{2}$.

 For ${\a\La_B \over M_6^4}<0$, all  allowed regions of the $(\r_m,w)$ plane are disconnected 
 (see Fig.~\ref{regionsL}d). In region I the system  evolves towards $\r_m \to 0^+$ and $w \to + \infty$. Its anisotropy increases 
   and tends in the asymptotics to $A \to \sqrt{2}$. 
      In region II, the system evolves towards $\r_m \to 0^-$ and $w \to - \infty$ 
       in the same way as in the $\La_B=0$ case. Its anisotropy decreases and in the 
        asymptotics it tends to $A \to \sqrt{2}$. In region III the system again 
         evolves to $w \to + \infty$ and $\a \r_m / M_{Pl}^2 \to -3 /  4$. 
          The anisotropy decreases,  but tends to a constant value $A > \sqrt{2}$.

The evolutions in the three regions resemble the ones plotted in 
Fig.~\ref{evol} for the $\La_B=0$ case. In all three regions the cosmological evolutions give rise to a regular geometry  at $r=0$ when the initial conditions are such that the system never crosses the lines of long dashing in Figs.~\ref{regionsL}a,b,d. For example,  this happens  for the evolutions in the region I for $\La_B>0$ and $w_s^-<-1<w<w_s^+$.

 From the above we conclude 
that, the relaxation of the tuning  between $w$ and $\r_m$, has as 
a consequence the appearance of new branches of brane world 
evolution (regions II, III), while in region I the system tends to 
the attractor fixed points, whenever they exist. For the cases 
when $\La_B>0$ the lines of isotropic tuning are attractors, with 
strength depending on the value of $\La_B$.

Furthermore, we observe that the anisotropy in all cases, apart 
the one of region I for $\La_B <0$, decreases much more quickly 
than in the four dimensional case, with the same initial conditions, 
but without the extra dimensional constraint. In region I for 
$\La_B <0$, the anisotropy increases and is larger than the one of the purely four 
dimensional case.

 \begin{figure}[t!] 
 
\begin{center} 
\epsfig{file=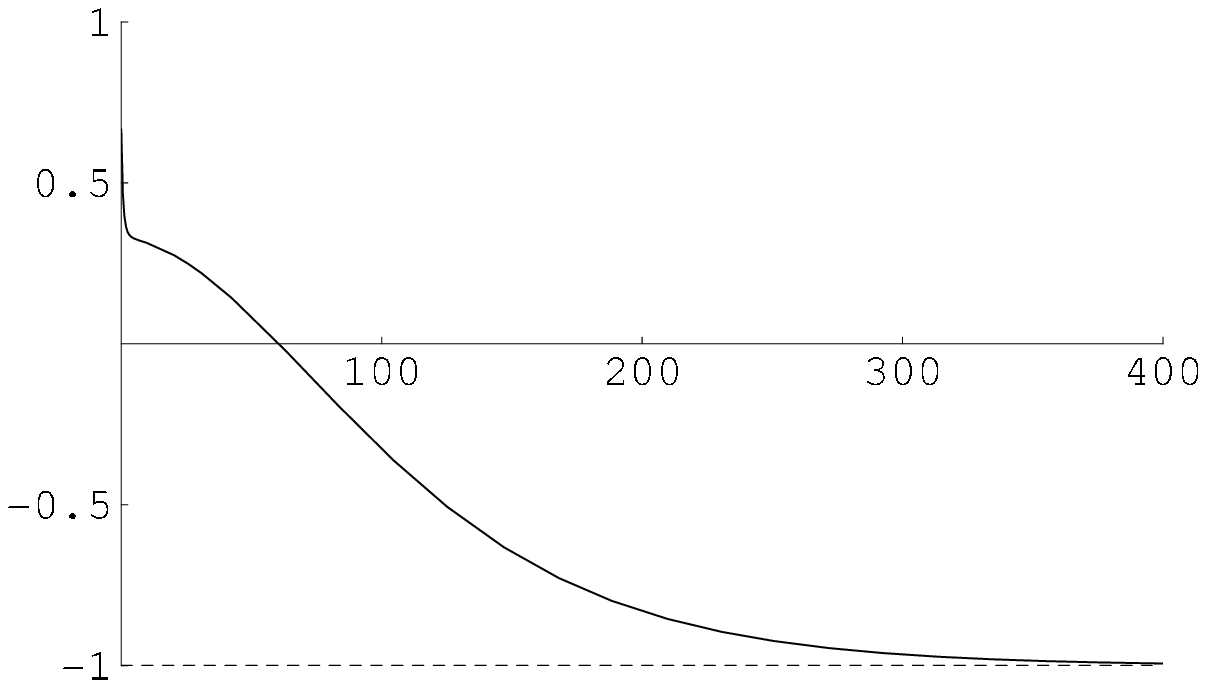,width=8cm,height=5cm} 
\epsfig{file=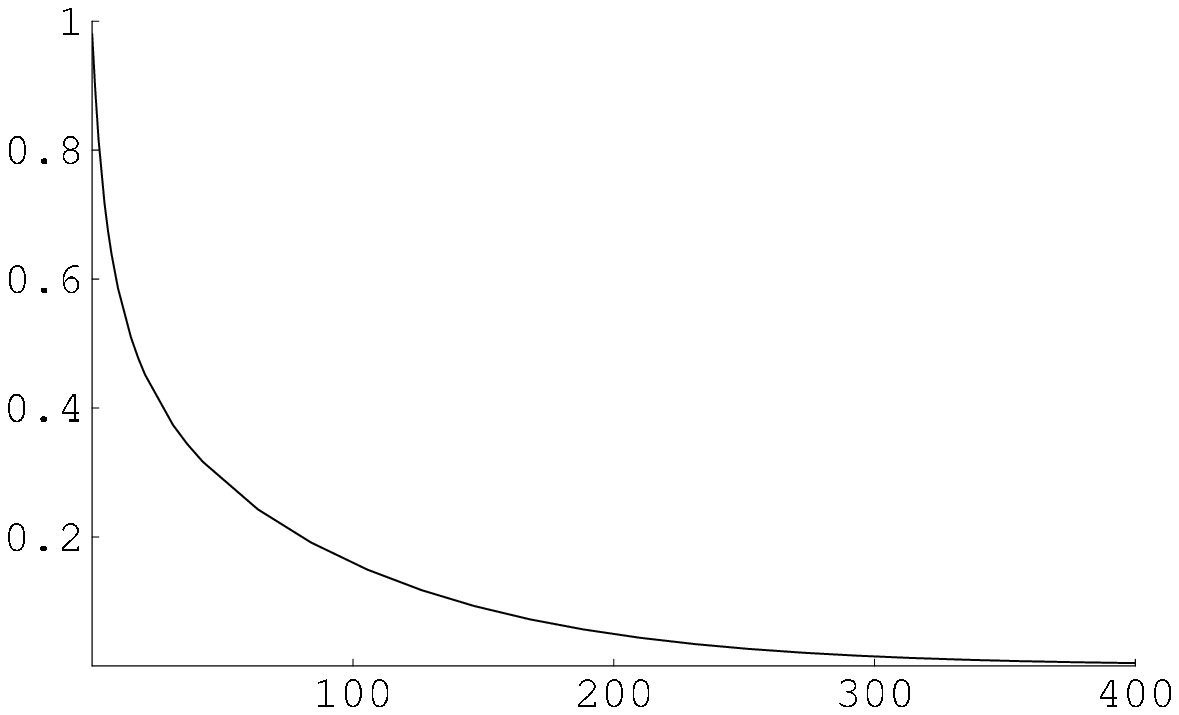,width=8cm,height=5cm} 
 
\begin{picture}(.01,.01)(0,0) 
 
\Text(200,35)[c]{$t$} 
\Text(-30,95)[c]{$t$} 
 
\Text(30,150)[c]{$A$} 
\Text(-200,150)[c]{$w$}

\end{picture} 
 
\vskip-5mm 
 
\caption{The  evolution of the equation of state $w$ and the anisotropy $A$ for the case of $\a \La_B / M_6^4 = 10^{-4}$ with initial condition $w_0=2/3$.  Until the fixed point is approached ($w \sim -1$), we have significant anisotropy $A \sim {\mathcal O}(1)-{\mathcal O}(10^{-1})$.} 
 
\label{smallLB} 
\end{center} 
\end{figure}

 Let us now examine again  the interesting possibility of 
\mbox{$0<\a\La_B / M_6^4 \ll 1$} with the transition between 
$w=1/3$ to  $w=0$ and finally to $w=-1$. As it can be inferred 
from the previous discussion, the system tracks the line of 
isotropic tuning and evolves towards the attractor fixed point 
with $w=-1$ (see Fig.~\ref{smallLB}). However, due to the small 
value of $\a\La_B / M_6^4$, the line of isotropic tuning is a {\it 
very weak attractor}. Most of the evolution is rather anisotropic 
with $A \sim {\mathcal O}(1)-{\mathcal O}(10^{-1})$  until  the 
fixed point is approached, in which region it drops to zero.

This large anisotropy makes the cosmological evolution 
phenomenologically problematic. In  order that the anisotropy is
acceptably small, the initial conditions  for the energy density 
and the equation of state should be {\it fine tuned} to lie very 
close to the line of isotropic tuning initially. 
 
In conclusion, by analyzing the anisotropic dynamics of the system 
we have seen that the lines of isotropic tuning are attractors for $\La_B>0$, 
with $\La_B$-dependent attracting strength. The most 
phenomenologically accepted evolutions with  $0<\a\La_B / M_6^4 
\ll 1$ do not isotropise quickly enough and thus  need a fine 
tuning in order to evolve with acceptable anisotropy.

\section{Conclusions} 
 
In this work, we studied the cosmological dynamics of a conical 
codimension-2 brane-worlds. We considered a theory with a 
Gauss-Bonnet term in the six-dimensional bulk and an induced 
gravity term on the three-brane. For simplicity, we considered 
that the bulk matter consists only of a cosmological constant 
$\La_B$ but the brane matter is general and isotropic. We then 
analyzed in detail the Einstein equations evaluated on the 
boundary.

We studied the system first for an isotropic metric ansatz. As was 
noted in \cite{Janpaper}, in the pure induced gravity dynamics 
there is a tuning between the matter allowed on the brane and in 
the bulk. This tuning, when  the matter in the bulk is only a 
cosmological constant, gives a precise relation between the matter 
density on the brane and its equation of state. If a Gauss-Bonnet 
term is added in the bulk, the constraint equation giving the 
previous tuning is modified by the addition of a Riemann squared 
term. However, since for isotropic evolutions the  Riemann tensor 
can be expressed in terms of the Ricci tensor and the scalar 
curvature, the conclusion about the presence of the tuning remains 
the same as in the induced gravity case. 
 
We therefore considered the combination of the two effects for the 
isotropic metric ansatz and made the further simplifying 
assumption that the brane  contains a tension contribution which 
gives a four-dimensional Einstein equation without a cosmological 
constant. The dynamics of the system of the four-dimensional 
equations plus the constraint equation of the extra dimensions, 
depend on the value of $\La_B$.  If $\La_B>0$ the fixed point has 
$w=-1$ and corresponds to a de Sitter vacuum. If $\La_B=0$, the 
system has a fixed point with equation of state $w=1/3$. Both of 
the previous fixed points are attractors. If $\La_B<0$ the system 
has no fixed point and the evolutions exhibit a runaway behaviour 
to $w \to +\infty$. 
 
The tight restriction on initial conditions on matter on the brane 
that the constraint equation imposes in the isotropic case, 
motivated us to look at the case where the initial ansatz is 
anisotropic. In this way, we could determine whether the above 
tuning corresponds to an attractor. For this purpose, we considered
 a particular kind of anisotropy. Since the 
Riemann tensor in this case  cannot be expressed in terms of the 
Ricci tensor and the scalar curvature, the previous studied 
constraint becomes now a dynamical equation for the anisotropy of 
the brane-world volume. The matter on the brane and its equation 
of state,  need not lie on the line of isotropic tuning, but can 
have values in vast regions of parametric space which was 
previously forbidden. There are always three distinct regions of 
parametric space, one of which has as a boundary the line of 
isotropic tuning. 
 
The dynamics of the system of the four-dimensional equations plus 
the extra dimensional equation, depend again on the value of 
$\La_B$. If the system starts its evolution in the region of 
parametric space which has as a boundary the line of isotropic 
tuning, it  tracks the latter line and isotropises towards the 
attractor fixed point with $w=-1$ for $\La_B>0$, or the one with 
$w=1/3$ for  $\La_B=0$. In the two other regions the system has a 
runaway behaviour $w \to\infty$ and does not isotropise. For 
$\La_B<0$, the system has always a runaway behaviour.  The 
important result of this analysis is that the line of isotropic 
tuning, coming from the constraints found in \cite{Janpaper},  is 
an attractor. However, for values of $\La_B$ which give acceptable 
cosmological evolutions, a fine tuning is unavoidable because of 
the weak strength of the above-mentioned attractor.

The above conclusions rely on the specific assumptions that we 
have made in order to simplify the dynamics of the system and 
might be altered if we change them. Let us remind the reader once 
again what has been assumed in the above analysis which may affect 
our final conclusions. Firstly, we consider only conical branes 
and in order to study cosmology on them we consider certain 
corrections to the gravity action (induced curvature and 
Gauss-Bonnet term). Secondly, we assume that the bulk matter has 
only the form of a cosmological constant. Thirdly, in studying the 
cosmological dynamics we consider that the tension contribution of 
the brane matter cancels the one induced by the deficit angle. 
 Finally, there is an implicit assumption in 
this work that the complicated bulk equations of motion, when 
integrated away from the brane, are well behaved (\eg no 
singularities).

Understanding the dynamics of codimension-2 branes is interesting 
because of the improvement they may make to the cosmological constant 
problem. There have been several models of self tuning relying on 
the properties of codimension-2 branes, where the vacuum energy of 
the brane does not curve the brane world-volume without fine 
tuning. It would be interesting to see what the cosmology of these 
models would be by repeating the above analysis for a bulk content 
which is provided by these models.

\section*{Acknowledgments} 
 
 A.P. wishes to thank the Physics Department of NTUA 
 for hospitality during the inception of this work.  This work 
is co-funded by the European Social Fund (75\%) and National 
Resources (25\%)-(EPEAEK II)-PYTHAGORAS.

\end{document}